# Effect of machine arithmetic errors for multi-turn invariant curve destruction

David Nikitin[*], Nataliya Stankevich[**]
*HSE University, Nizhny Novgorod, Russia*
[*]dsnikitin_1@edu.hse.ru, [**]stankevichnv@mail.ru

*Using the example of three-dimensional Mira map, it is shown that the destruction of a multi-turn invariant curve can occur through the appearance of local multiple bends. It was found that, depending on the precision of machine arithmetic, a complication of the multi-turn invariant curve can be observed, which is a numerical artifact. Artifact can be avoided with increasing of machine arithmetic accuracy.*

Keywords: torus doublings; machine arithmetic error; invariant curve

**Introduction**

Period-doubling bifurcation is one of the fundamental bifurcations that can occur with a stable fixed point in a map and with a limit cycle in a flow system (in the Poincaré section, it also corresponds to a fixed point) [1-4]. In this case, the multiplier of the point becomes equal to -1 ($\mu$ = -1); the point loses stability and becomes unstable or becomes a saddle, while a period-2 cycle emerges in its vicinity. A cascade of such bifurcations leads to the emergence of a set of cycles of period $2^n$, where *n* is the number of bifurcations. The infinite cascade results in the onset of chaos, which is referred to as the Feigenbaum scenario for the transition to chaos [5].

Period-doubling bifurcations can also undergo more complex objects, such as invariant curves in discrete systems or tori in flow systems [6-11]. This bifurcation likewise corresponds to the loss of stability of an invariant curve when its multiplier crosses -1: the invariant curve becomes saddle-type, and a doubled invariant curve is born in its vicinity. The complexity of the base objects (tori, doubled tori, etc.) makes the analysis of such bifurcations considerably challenging, and the theory of bifurcations of invariant curves currently has a number of open questions. These include, for example, the finiteness of cascades of torus-doubling bifurcations [8, 10]. The literature provides numerous examples where a finite number of period-doubling bifurcations of invariant curves (tori) is observed, after which the torus becomes resonant, loses smoothness, and breaks down [8, 11]. The preservation of an invariant curve indeed requires specific conditions (the frequencies must maintain an irrational relationship), which are easily satisfied when the frequencies are fixed, as in the case of an external quasi-periodic forcing. If the system parameters vary and are not strictly fixed, the emergence of resonances, leading to the loss of smoothness of the invariant curve, is the most commonly observed outcome. However, several studies present examples of extended cascades of period-doubling bifurcations of invariant curves [12-13], offering some hope for the existence of infinite cascades. Furthermore, a number of works discuss the swollen shape type bifurcations, where period-doublings of invariant curves transform into local swellings [14-18].

During a period-doubling bifurcation of a torus or an invariant curve, the attractor in the system becomes increasingly complex very rapidly, rendering its analysis difficult. Following successive doublings, invariant curves may exhibit local clustering of points and may pass in very close proximity to one another. Furthermore, when resonances emerge on multi-turn invariant curves, their periods can become extremely large, which also presents significant challenges for analysis. Moreover, when investigating such subtle effects, influences related to numerical approximation errors and machine precision begin to manifest. In this paper, we aim to demonstrate the manifestation of machine precision effects during the destruction of a multi-turn invariant curve, using the example of a period-doubling cascade of an invariant curve in the three-dimensional generalized Mira map.

## 2. Object and Methods of Study

As the object of study, we will consider a three-dimensional diffeomorphism, namely the three-dimensional generalized Mira map:

$$x_{n+1} = y_n,$$
$$y_{n+1} = z_n, \quad (1)$$
$$z_{n+1} = Bx_n + Az_n - Cy_n - y_n^2.$$

This map is extensively discussed and analyzed in [19]; in particular, this reference provides a detailed discussion of the origin of the map's name, as well as its relationship to the class of Hénon maps and to the Lomelí maps. Map (1) features three dynamical variables ($x$, $y$, $z$) and three control parameters ($A$, $B$, $C$). Depending on these parameters, the system can exhibit various types of bifurcations and different dynamical regimes, including quasi-periodic and chaotic behavior.

Since the primary aim of this work is to investigate period-doublings of invariant curves and their subsequent destruction, we will employ the method of Lyapunov exponents charts to localize the structures of interest in the parameter space. The method is as follows: the parameter plane is scanned with a given step size, and for each pair of parameters, the full spectrum of Lyapunov exponents is computed [20]. Based on this spectrum, the type of dynamical regime is classified. For a three-dimensional diffeomorphism, six types of dynamical regimes can be distinguished; these are listed in Table 1.

Table 1. Correspondence between the types of dynamical regimes, the signature of the Lyapunov exponents, and the notation and colors used in the Lyapunov exponent charts

| Type of dynamical regime | Signature of the Lyapunov exponents spectrum ($\Lambda_1, \Lambda_2, \Lambda_3$) | Label | Color |
| --- | --- | --- | --- |
| Fixed point, stable cycles | (-, -, -) | FP, CP | blue |
| Invariant curve | (0, -, -) | IC | green |
| 2-torus, line of invariant curve doubling | (0, 0, -) | T2, ICD | yellow |
| Chaos | (+, -, -) | C | red |
| Chaos with zero Lyapunov exponent | (+, 0, -) | C0 | pink |
| Hyperchaos | (+, +, -) | HC | aqua |

Furthermore, the Lyapunov exponent spectrum allows for the detection of period-doubling bifurcation lines of invariant curves [21-22]. At the moment of a doubling bifurcation, the largest Lyapunov exponent is zero, the second exponent approaches zero, and after the bifurcation, it becomes negative again. Given that a threshold defining zero is used when constructing the chart, lines where two exponents become zero appear within the regions corresponding to invariant curves; these lines correspond to torus-doubling bifurcation curves. For our Lyapunov exponent charts, the zero threshold was set to $10^{-3}$.

Figure 1 presents a Lyapunov exponent chart on the parameter plane ($A$, $C$) for $B = 0.5$. In papers [19, 23], it was shown that cascades of period-doubling bifurcations of invariant curves are observed in the vicinity of a codimension-2 point. It is precisely the neighborhood of this region that is depicted in Fig. 1. A characteristic feature of these regions is the emergence of areas of chaos with one positive and one zero Lyapunov exponent [24].

The chart clearly shows that the fixed point undergoes a Neimark–Sacker bifurcation, resulting in the birth of an invariant curve (green region) in the system. Within the green region, yellow lines corresponding to period-doublings of the invariant curve are observed. It is evident that for larger values of the parameter $A$, these doublings lead to the emergence of chaos with one zero Lyapunov exponent. Within this region, a quasi-periodic shrimp-shaped structure is also observed [24]. As the parameter $A$ decreases, the quasi-periodic structures within the chaos break down, and chaos with one zero exponent transforms into chaos or hyperchaos; regions of periodic regimes become visible on the map. It is in this area, where resonances of large period appear, that we will examine the features of the period-doubling cascade of the invariant curve and its destruction. In Fig. 1, the dashed line indicates the route along which we will conduct a detailed analysis in the following sections.

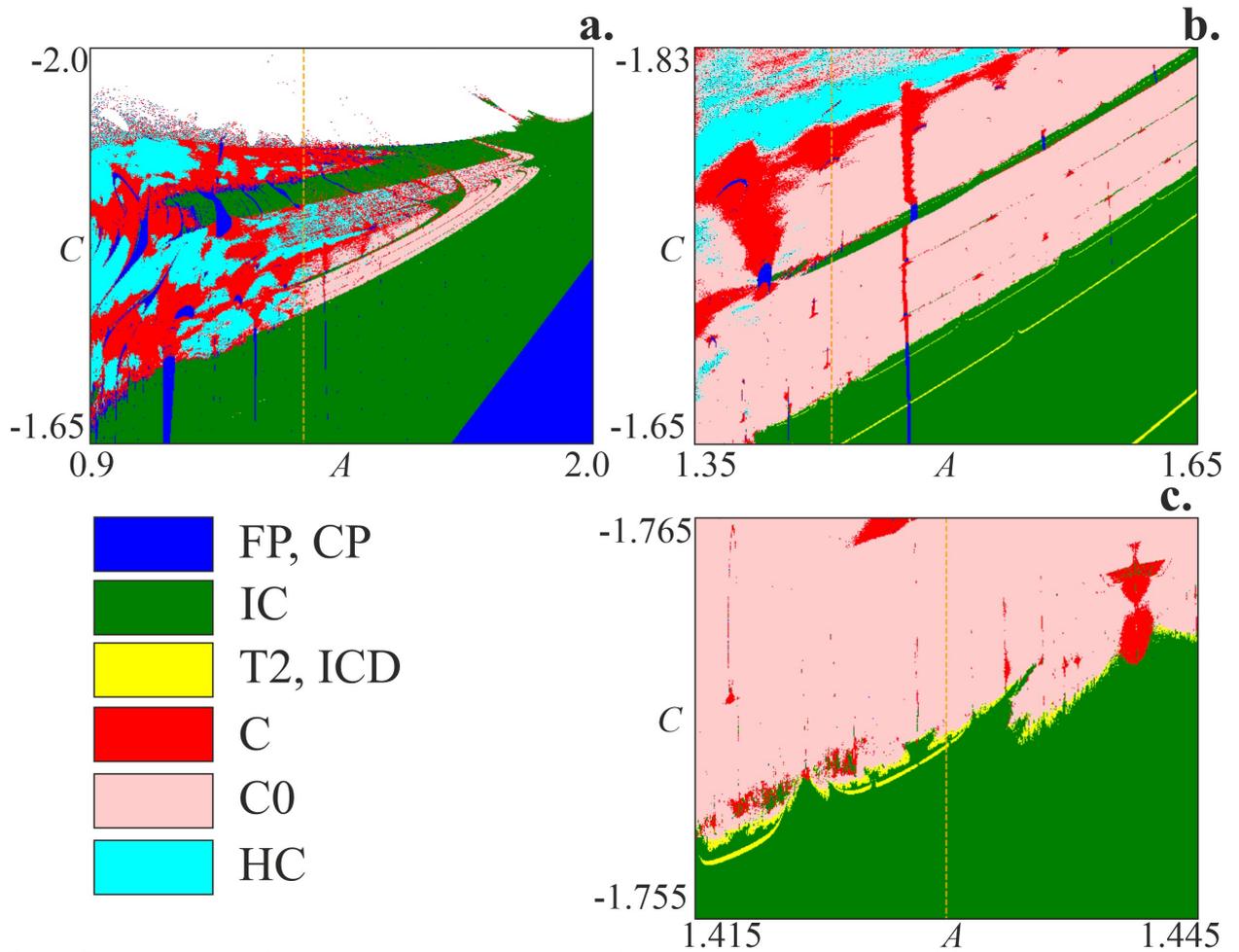

Fig. 1. Fragments of the Lyapunov exponent chart for the three-dimensional generalized Mira map (1) in the vicinity of the region where period-doubling bifurcations of invariant curves and the transition to chaos are observed, $B=0.5$

We now proceed to investigate the region where the transition to chaos is observed, located farthest from the codimension-2 point. Figures 1b and 1c show two zoomed fragments of these regions. In the chart in Fig. 1b, three period-doubling bifurcation lines of the invariant curve are clearly discernible. The enlarged fragment in Fig. 1c demonstrates that along the line $A=1.43$, three period-doubling bifurcations also occur, after which the invariant curve becomes chaotic.

## 3. Period-Doubling Cascade of the Invariant Curve and Its Destruction

For a more detailed analysis of the destruction and chaotization scenario of the invariant curve, we employ a one-parameter bifurcation analysis. Figure 2a presents the graphs of the two largest Lyapunov exponents as functions of the parameter $C$ for $A=1.43$ (orange dashed line in Fig.1). The Neimark–Sacker bifurcation (*NS*) leading to the birth of an invariant curve is clearly visible, followed by two period-doubling bifurcations of the invariant curve (*ICD$_1$*, *ICD$_2$*) and its subsequent destruction, marking the transition to chaos.

Figures 2b–2e illustrate the corresponding transformations of the phase portraits at these bifurcations: initially, the invariant curve becomes two-turn (Fig. 2b), then four-turn (Fig. 2c). An 8-turn invariant curve is also depicted (Fig. 2d). As noted above, following the third period-doubling bifurcation, the 8-turn invariant curve exhibits a structure where its two components are located in such close proximity that they are difficult to distinguish. Figure 2e shows an example

of a chaotic attractor observed at $C=-1.76$. Its structure retains the shape of the 4-turn invariant curve.

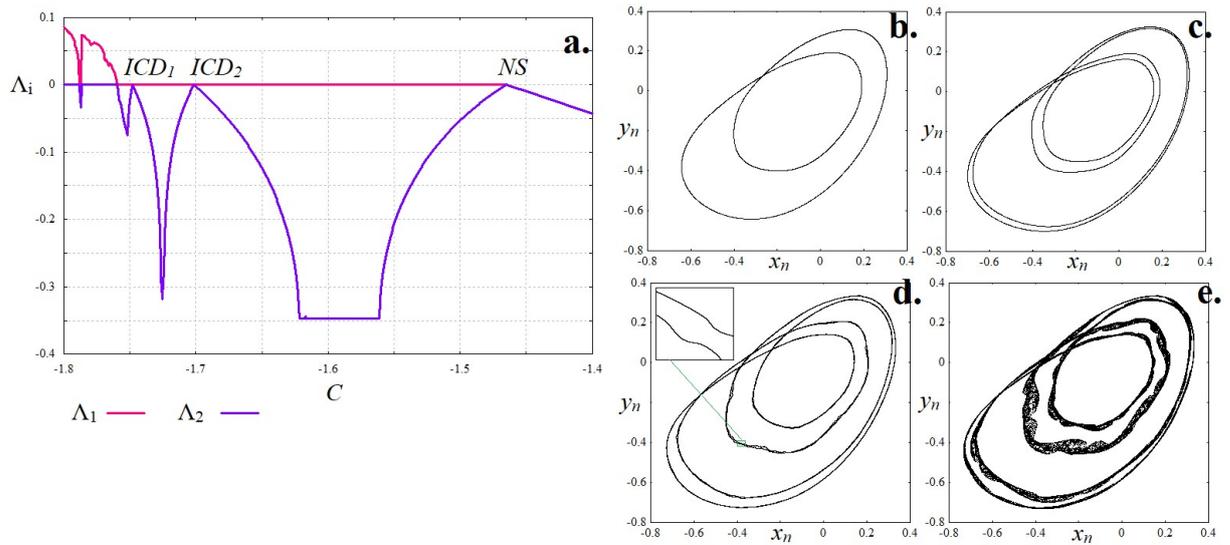

Fig. 2. Cascade of three period-doubling bifurcations of the invariant curve and the transition to chaos in the three-dimensional generalized Mira map (1) for $B=0.5$, $A=1.43$. **a.** Graphs of the two largest Lyapunov exponents; **b.–e.** Two-dimensional projections of the phase portraits for: **b.** $C=-1.73$; **c.** $C=-1.75$; **d.** $C=-1.7594$; **e.** $C=-1.76$

To understand how the final destruction of the 8-turn invariant curve occurs, we examine in more detail the transformations of the phase portraits and the dynamics of the Lyapunov exponents in the vicinity of the onset of chaos. Figure 3 presents fragments of the Lyapunov exponent graphs, along with enlarged fragments of the phase portraits illustrating the destruction of the invariant curve.

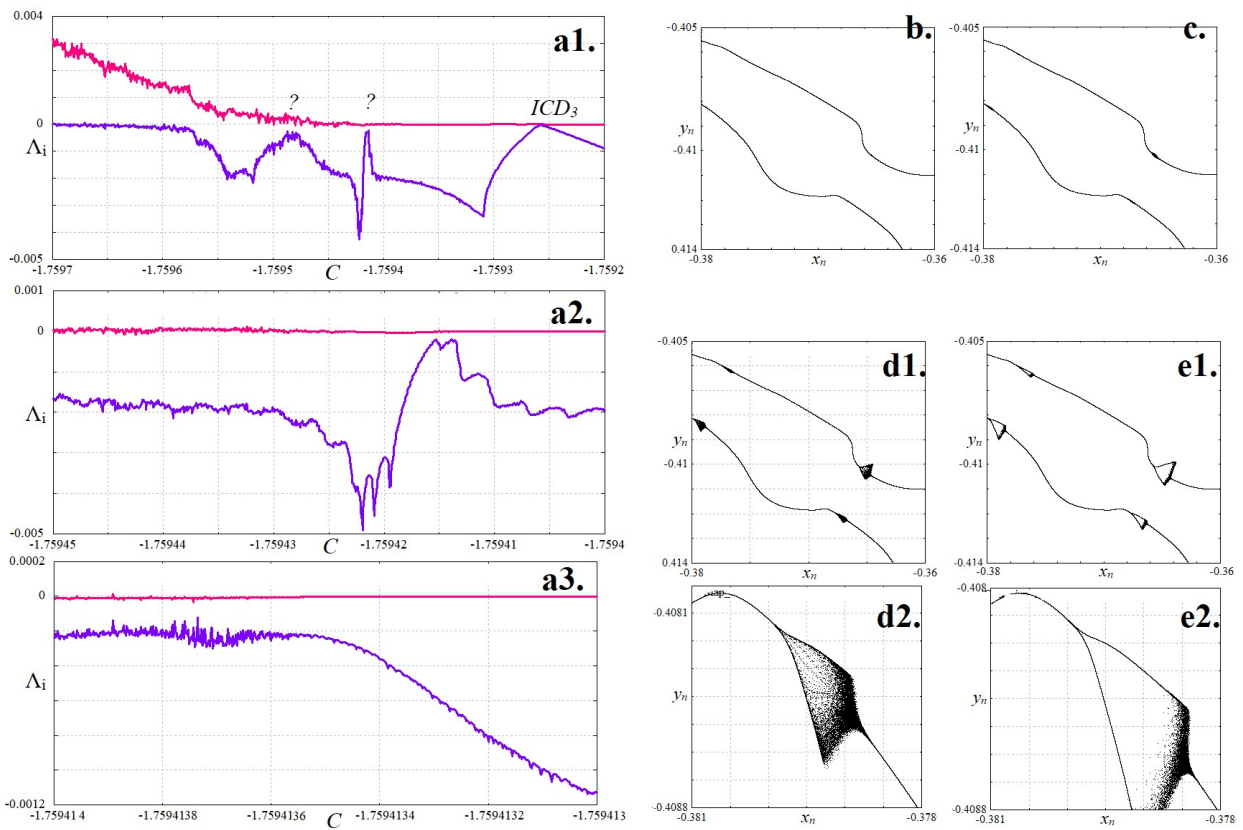

Fig. 3. Destruction of the 8-turn invariant curve in the three-dimensional generalized Mira map (1) for $B=0.5$, $A=1.43$. **a.** Enlarged fragment of the graphs of the two largest Lyapunov exponents; **b.–e.** Two-dimensional projections of the phase portraits and their enlarged fragments for: **b.** $C=-1.75946$; **c.** $C=-1.75947$; **d.** $C=-1.7595$; **e.** $C=-1.7596$

In the enlarged fragment of the Lyapunov exponent graph, the point corresponding to the third period-doubling bifurcation of the invariant curve ($ICD_3$) is clearly identifiable. As parameter $C$ is further decreased, a variation in the second Lyapunov exponent resembling another doubling bifurcation is observed: the second exponent approaches zero. However, the magnified view clearly shows that it does not reach zero and becomes negative again. It is noteworthy that after passing this point, small fluctuations around zero appear in the graph of the largest Lyapunov exponent. With a continued decrease in $C$, another approach of the second Lyapunov exponent towards zero can be seen. Concurrently, the fluctuations of the largest exponent become increasingly pronounced, and the Lyapunov exponent graph allows for the unambiguous classification of the behavior as chaotic. It is evident that the chaos initially appears in its classical form, characterized by a positive largest Lyapunov exponent and a negative second exponent. As $C$ increases, this regime transforms into chaos with one zero Lyapunov exponent.

Figures 3b–3e show enlarged fragments of the projections of the phase portraits as the parameter $C$ is varied through the region where the second Lyapunov exponent increases. It is clearly observed that small local swellings appear on the smooth invariant curve, which grow as the parameter $C$ decreases. In the further enlarged fragments, it is evident that these swellings are shaped like locally bifurcated invariant curves, which are filled with phase trajectories from the inside.

## 4. Influence of Machine Precision on the Destruction of Invariant Curves

Machine arithmetics, which is associated with the use of computing devices, has features related to the representation of numbers and the execution of arithmetic operations. These features relate to the operating principles, number systems and arithmetic, as well as errors that occur during calculations. When investigating such delicate objects as those presented in Fig. 3, it is necessary to consider that, within the regions of local swellings, the machine precision of standard programming packages may be insufficient.

For the calculations of the illustrations shown in Figs. 1–3, we used programs prepared in Python, C++, and Pascal. For all three programming languages, we employed double-precision arithmetic, which corresponds to 15–19 decimal digits. This level of precision can be readily verified and increased using a specialized library in Python like Mpmath. Using this library, we tested the reproducibility of the local swellings on the multi-turn invariant curves.

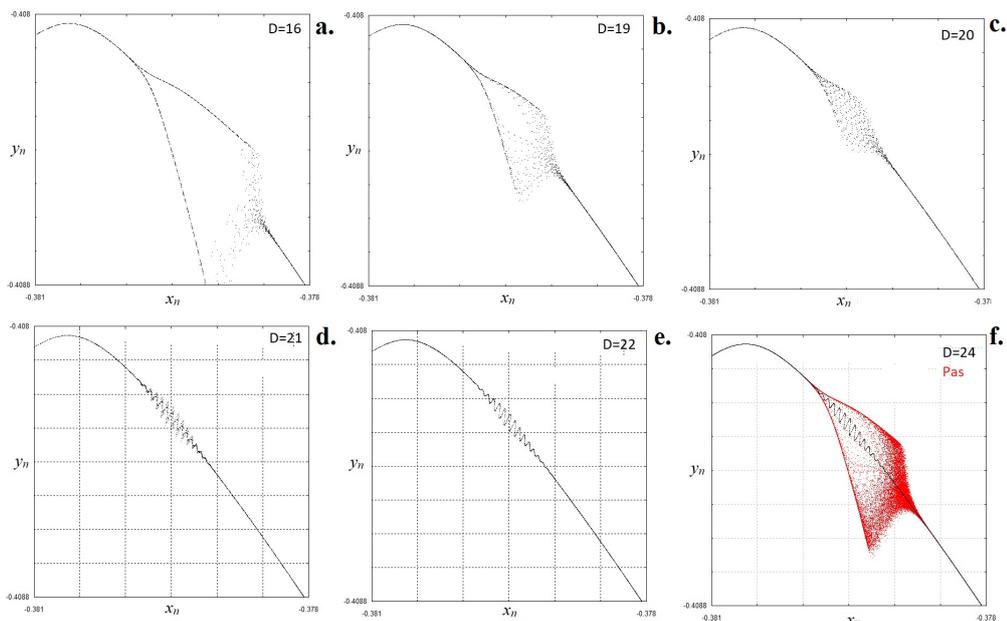

Fig. 4. Testing the precision of machine computation using enlarged fragments of two-dimensional projections of phase portraits for map (1) at $B=0.5$, $A=1.43$, $C=-1.7594137$. D is number of decimal places

Figure 4 presents examples of fragments of the phase portrait projection for the case corresponding to Fig. 3d. It is clearly observed that increasing the machine precision of the calculation transforms the bifurcation of the invariant curve into local fluctuations. When accounting for 22 decimal digits, the invariant curve becomes uniquely defined. Further increasing the precision no longer alters it.

Figure 5 presents illustrations of precision testing for $C=-1.759414$. It is clearly observed that in this case, increasing the precision also alters the phase portrait; however, the splitting of the invariant curve persists.

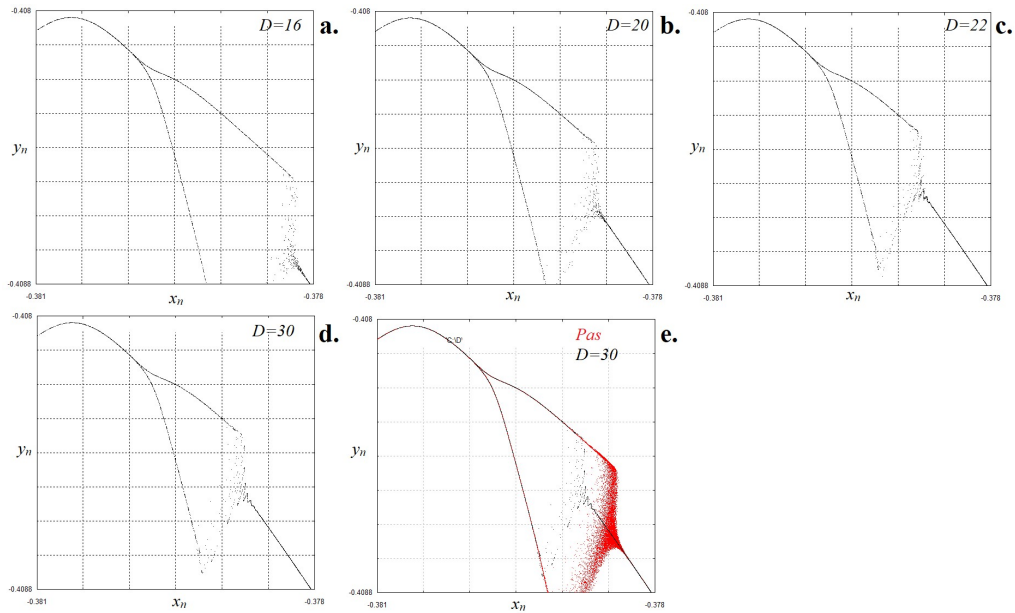

Fig. 5. Testing the precision of machine computation using enlarged fragments of two-dimensional projections of phase portraits for map (1) at $B=0.5$, $A=1.43$, $C=-1.759414$. D is number of decimal places

## Conclusions

In this work, we have investigated the dynamics of period-doubling cascades of invariant curves and their subsequent destruction in the three-dimensional generalized Mira map. Using Lyapunov exponent charts, we identified regions in the parameter plane where invariant curves undergo a sequence of period-doubling bifurcations, ultimately leading to chaos. A detailed one-parametric analysis revealed that after three period-doubling bifurcations, the invariant curve becomes 8-turn and exhibits local swellings that precede its chaotization.

The main focus of the study is the significant influence of machine precision on the numerical observation of the delicate structures of multi-turn invariant curves. By employing the Mpmath library in Python to increase computational accuracy, we demonstrated that the apparent bifurcations and swellings of the invariant curve can be artifacts of insufficient precision. Specifically, at certain parameter values, increasing the number of decimal places transformed the observed bifurcations into smooth fluctuations of the invariant curve. At other parameter values, however, the bifurcation persisted even at higher precision, indicating a dynamical feature.